\documentclass[aoas,nameyear,seceqn,dvips]{arximspdf}
\usepackage{graphics}

\doi{10.1214/07-AOAS150}
\volume{2}
\issue{2}
\pubyear{2008}
\firstpage{536}
\lastpage{549}

\begin{document}
\begin{frontmatter}

\title{Should the democrats move to the left on economic policy?\protect\thanksref{T1}}
\runtitle{Should the democrats move to the left on economic policy?}
\thankstext{T1}{Supported by U.S. National Science
Foundation and the Columbia University Applied Statistics Center.}

\begin{aug}
\author[A]{\fnms{Andrew} \snm{Gelman}\corref{}\ead[label=e1]{gelman@stat.columbia.edu}}\and
\author[A]{\fnms{Cexun Jeffrey} \snm{Cai}\ead[label=e2]{jeffccx@gmail.com}}
\runauthor{A. Gelman and C. J. Cai}
\affiliation{Columbia University}
\address[A]{Department of Statistics \\
and \\
Department of Political Science\\
Columbia University\\
New York\\
USA\\
\printead{e1}\\
\phantom{E-mail:} \printead*{e2}} 
\end{aug}

\received{\smonth{11} \syear{2007}}
\revised{\smonth{11} \syear{2007}}

%
\begin{abstract}
Could John Kerry have gained votes in the 2004 Presidential election
by more clearly
distinguishing himself from George Bush on economic policy? At first
thought, the
logic of political preferences would suggest not: the Republicans are
to the right
of most Americans on economic policy, and so in a one-dimensional space
with party positions
measured with no error, the optimal strategy for the Democrats would be
to stand
infinitesimally to the left of the Republicans. The median voter
theorem suggests
that each party should keep its policy positions just barely
distinguishable from the opposition.

In a multidimensional setting, however, or when voters vary in their
perceptions of the
parties' positions, a party can benefit from putting some daylight
between itself and
the other party on an issue where it has a public-opinion advantage
(such as economic
policy for the Democrats). We set up a plausible theoretical model in
which the
Democrats could achieve a net gain in votes by moving to the left on
economic policy,
given the parties' positions on a range of issue dimensions. We then evaluate
this model based on survey data on voters' perceptions of their own positions
and those of the candidates in 2004.

Under our model, it turns out to be optimal for the Democrats to move
slightly to
the \textit{right} but staying clearly to the left of the Republicans'
current position on economic issues.
\end{abstract}

%
\begin{keyword}
\kwd{Median voter}
\kwd{Presidential election}
\kwd{public opinion}
\kwd{spatial model of voting}.
\end{keyword}

\end{frontmatter}

\setcounter{footnote}{1}
\section{Introduction}

In the 2004 presidential election campaign, it has been suggested
that voters saw little difference between the parties on economics but
large differences on other issues. The Democrats are traditionally
closer than the Republicans to the average voter's view on the economy.
Should the Democrats have moved to the left on economic issues? Could
such a strategy win them votes? We study this using a theoretical model
and survey data.

\subsection{Candidate positions and the median voter theorem}

\begin{figure}

\includegraphics{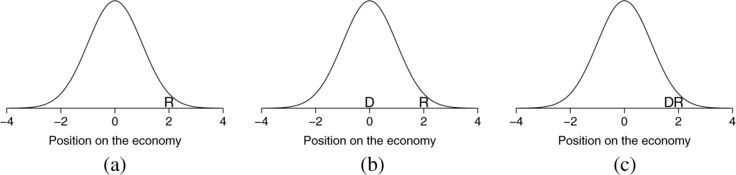}

\caption{Some possibilities in a one-dimensional spatial model: the
curve indicates the opinions of voters on economic issues, and D and R
show the positions of the Democratic and Republican parties,
respectively. In all three pictures the Republicans are right of
center. In \textup{(b)}, the Democrats are at the median voter; in \textup{(c)}, the
Democrats are just barely to the left of the Republicans, thus
optimizing their vote share if the Republicans are not free to move.
(We are assuming here that the Republican position is fixed, perhaps
because they are the incumbent party or perhaps because of strong
policy preferences.)\label{1d}}
\end{figure}

In a two-party system the median voter theorem states that it is in
each party's best interest to move toward the center (the median) of
the distribution of voters [Hotelling (\citeyear{hot1929}) and
Downs (\citeyear{dow1957})]. If either
party is not at the median, the other party has a winning strategy. For
example, in Figure \ref{1d}(a) the Republicans have a position to the
right of the average voter. If the Democrats sit at the median [see
Figure \ref{1d}(b)], they will attract more than half the voters. But the
Democrats will do even better by moving just infinitesimally to the
left of the Republicans [see Figure \ref{1d}(c)] and getting the votes of
everyone to the left.

This analysis ignores the possibility that the Republicans can also
move (an issue to which we return in Section \ref{practical}). If both
parties are free to move to optimize their votes, they will converge to
an equilibrium where they are both at the median.

The median voter theorem is regularly falsified by actual data.
Politicians regularly depart from the median [Poole and Rosenthal (\citeyear{pooRos1997})]
despite there being clear evidence of an electoral benefit for
having moderate positions [Gelman and Katz (\citeyear{gelKat2005})]. Legislators'
distances from the median have been found to be correlated with
district characteristics [Gerber and Lewis (\citeyear{gerLew2004})]. There are many
practical reasons for politicians to move away from the center.
Ideological positioning is only one of the factors influencing
election outcomes, and a candidate might well, for example, sacrifice
an estimated 2\% of the vote in order to be better positioned to
implement desired policies in the event of an election victory. There
are also other constituencies to satisfy (including campaign
contributors, party activists and primary election voters). We are
assuming that ideological stances reflect real policy issues---or, to
put it another way, we are assuming that the candidates have already
performed whatever ideological posturing they can, and that changes in
their spatial ``locations'' can be effected only by changes in policy positions.

The median voter theorem also becomes more complicated with constraints
on candidate positions, multiple issue dimensions, and variation among
voters in perceptions of candidates. These are the directions we
explore in this paper, to see whether the Democrats might gain from
moving to the left on economic issues, apparently contradicting the
one-dimensional picture in Figure \ref{1d}.

\section{Simple theoretical models}\label{theory}

We shall illustrate the potential benefits for the Democrats to move
using a simple spatial voting model with error [following Erikson and
Romero (\citeyear{eriRom1990})] in one, two and three dimensions. In each model we set
up a simple unimodal distribution for voter preferences, place the two
parties in this distribution, and then consider what happens to the
Democrats' share of the vote if we change their position on the
economic dimension.

\subsection{Spatial voting models in \textup{1,} \textup{2} and \textup{3} dimensions}

\subsubsection*{One-dimensional model}

We stipulate that voters' individual positions on the economy follow a
unit normal distribution, with negative and positive values being
liberal and conservative. We further assume that the Republicans'
position is $+$2 (very conservative) and that the Democrats start at $+$1
(somewhat conservative). If we now let the Democrats move freely, it is
clear that their optimal position is around $+$1.9999, so that they will
get all the votes of the people to their left. [See Figure \ref{1d}(c).]
This is the sort of reasoning that leads the Democrats to move as close
as possible to the Republicans while staying just slightly toward the center.

This model also predicts that the Democrats will get over 90\% of the
vote! In actual elections, though, the Republicans actually do pretty
well, but maybe not specifically because of their conservative economic
policies on issues such as tax rates, trade, the minimum wage, and so forth.

\subsubsection*{Two-dimensional model}

\begin{figure}

\includegraphics{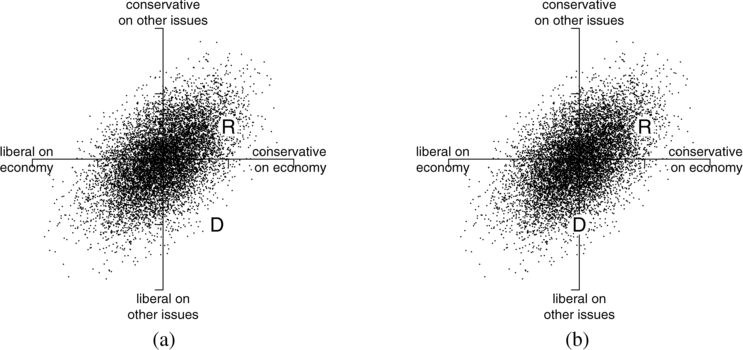}

\caption{Some possibilities in a two-dimensional spatial model: the
points in the scatterplot indicate the opinions of voters on economic
and other issues,
and D and R show the positions of the Democratic and Republican
parties, respectively. In both pictures the Republicans are right of
center on economics and on other issues, and the Democrats are left of
center on other issues. In \textup{(a)}, the Democrats are at just barely to the
left of the Republicans on economics; in \textup{(b)}, the Democrats are at the
median. Unlike in the one-dimensional scenario (see Figure \protect\ref{1d}),
the Democrats are better off separating themselves from the Republicans
on economic issues.\label{2d}}
\end{figure}

We now move to a two-dimensional model, whose dimensions we label as
``economic issues'' and ``all other issues.'' Figure \ref{2d}(a) shows
our assumptions:
the voters have a bivariate normal distribution with correlation~0.5
(fiscal conservatives are commonly, but not always, social
conservatives), the Republicans are at $(+2,+1)$---very conservative in
economic policy, somewhat conservative otherwise---and the Democrats
are at $(+1,-2)$---moderately conservative economically, very liberal
otherwise. Finally, we assume the two dimensions are equally important
and that a voter will prefer the candidate who is closer (in Euclidean
distance).

In Figure \ref{2d}(a) more voters are closer to the Republicans'
position than to the Democrats'. Although the Democrats are slightly
more moderate on economic issues, they are further from the majority of
the voters.

Now suppose the Democrats have the freedom to alter their
position---but only on the economic dimension (see Section \ref
{practical} for discussion of this point). Should they move leftward
(toward the median voter) or rightward (toward the Republicans, in the
way that would be recommended from the one-dimensional model)? The
answer is: unlike in one dimension, the Democrats should move to the
left! Figure~\ref{2d}(b) shows that if the Democrats move to $(0,-2)$,
they pick up votes from the Republicans.

\begin{figure}[b]

\includegraphics{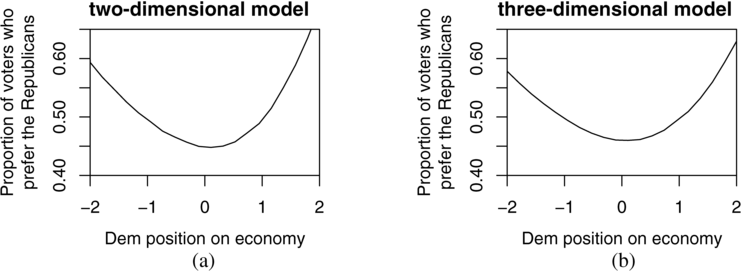}

\caption{\textup{(a)} Proportion of voters who would prefer the Democrats, as a
function of the party\textup{'}s position on the economy, assuming that the
Democrats\textup{'} positions on other issues is fixed at $-2$ and that the
Republicans\textup{'} positions are fixed at $+$2 on the economy and $+$1 on other
issues; see Figure~\protect\ref{2d}. Under these conditions, the Democrats are
best off being very slightly to the right of center. They should not be
at $+$1.999 as would be implied by the simple one-dimensional spatial
theory. \textup{(b)} Proportion of voters who would prefer the Democrats, as a function
of the party\textup{'}s position on the economy, in a similar three-dimensional
spatial model. As in two dimensions, it benefits the Democrats to
clearly distinguish themselves from the Republicans on the
economy.\label{curves}}
\end{figure}

More generally, Figure \ref{curves}(a) (computed by simulation using
10,000 voters randomly-sampled from the bivariate normal distribution)
shows the proportion of voters who would prefer the Republicans, under
the spatial model, as a function of the Democrats' position. In this
configuration the Democrats benefit by being slightly more
conservative than the average voter but still clearly separated from
the Republicans on the economic dimension, so as to be closer to the
mass of voters in the two-dimensional space.

This model seems unrealistic, as it predicts that the Republicans
support could vary from the range of 45\% to 65\%. We shall discuss
more realistic models below. The point here is that even the simple
spatial model has interesting implications when moving beyond one
dimension, leading to a violation of the often-assumed rule that the
Democrats would gain by being as conservative as possible (and,
conversely, that the Republicans should be as liberal as possible).

\subsubsection*{Three-dimensional model}

Figure \ref{curves}(b) shows that a similar pattern holds for the
three-dimen\-sion\-al model. Here the dimensions are economic, foreign,
and social policy. We assume the voters follow a normal distribution
with correlation 0.5 among each pair of dimensions, with Republicans at
$(+2,+1,+1)$ and Democrats at $(x,-1,-2)$, where we consider values of
$x$ ranging from $-2$ to $+2$. If we consider $x=+1$ to be the status
quo, we see that, as in two dimensions, the Democrats would do better
to move to the left, toward the mass of the voters.

\subsection{Varying the model specification}

Our spatial model can be generalized in many ways, two of which we
consider here. First, we suppose that different voters have different
perceptions about where candidates stand on the issues. Second, we
suppose that preferences depend on factors other than ideology.

\subsubsection*{Differing perceptions of candidate issue stances}

Different voters have different views about where the candidates stand
on the issues. This variation can be expressed as an error term in our
model of candidate positions, and the distribution of these perceptions
can be estimated using survey data.

What will be the effect of adding uncertainty about party positions? In
the one-di\-men\-sion\-al model, it can make a big difference. Once we
add uncertainty, it is no longer optimal for the Democrats to be
infinitesimally to the left of the Republicans. Even in one dimension, it
makes sense for the Democrats to move to the left---that is, toward the
center---to establish a clear difference for the voters [Erikson and
Romero (\citeyear{eriRom1990})]. In addition, as discussed by
Chappell and Keech (\citeyear{chap1986}),
in the presence of uncertainty about party positions, it makes sense
for parties to move toward their policy preferences.

Even in the absence of motivation or turnout effects, once there is
uncertainty or variation in perceptions of candidates, a party can gain
by clearly delineating itself on issues for which it has popular
support (such as the economy for Democrats). Separation is beneficial
in itself if it conveys the relative positions of the parties to more
of the voters.
We shall explore this further in our empirical analysis in the next section.

In two or more dimensions, adding uncertainty doesn't change the
fundamentals of the model: depending on the positions of the parties
and the distribution of the voters, it can still makes sense for
the Democrats to move toward the center, or to distance themselves from
the Republicans on economic issues.

\subsubsection*{Allowing preferences to depend on factors other than ideology}

There are also the ``valence issues.'' Suppose all the voters'
positions on issues are fixed, and the candidate positions are fixed.
Then the economy booms. This will benefit the party in power, even if
basic views on economy are not changed. A change in the economy might
also change voters' views about economic issues, but the ``valence''
idea is that, in addition to any such fundamental change, there will be
a shift in preferences. This would be expressed as an additive term in
the utility model. Thus, the relative utility of the Democrats,
compared to the Republicans, for voter~$i$, would be $\|x_i - R\|^2 -
\|x_i - D\|^2 + \mbox{shift}$, where $x_i$ is the (multidimensional)
ideological position of voter $i$, $R$ and $D$ are the positions of the
two parties, and the shift represents valence issues. ``Valence
issues'' in this definition also include incumbency advantage, unequal
spending, and any other advantages for one party or another, beyond
issue positions. This framework is consistent with the findings of
Rosenstone (\citeyear{ros1984}) and others that election outcomes are predictable
given measures of ideological difference and recent economic
conditions. As Groseclose (\citeyear{gro2001}) points out, a candidate who is weaker
on valence issues can be motivated to move away from the center on issues.

One can also alter the model in other ways. For example,
so far we have assumed a quadratic utility function---that is, based on
squared Euclidean distance between candidates and voters. Instead we
can define utility based on absolute-value distance [i.e., changing
from $d(x,y)=\sum_j(x_j-y_j)^2$ to $d(x,y)=\sum_j|x_j-y_j|$, in both
cases summing over dimensions $j$]. Changing this distance function has
little effect on the basic patterns we have found.
The utility function can also be generalized so that some issues are
more important than others---that is, a weighted sum over dimensions
instead of a simple sum.

\section{Empirical data on voter and candidate positions on issues}

\begin{figure}

\includegraphics{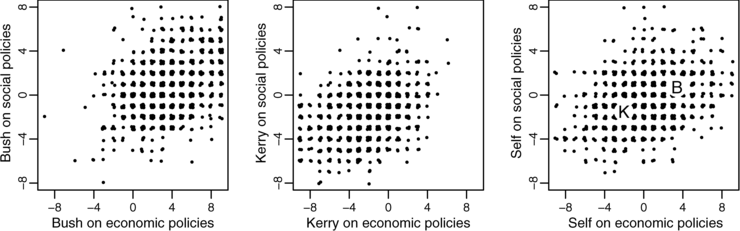}

\caption{Respondents\textup{'} views of Bush, Kerry, and themselves on a scale
of $-9$ (extremely liberal) to $+9$ (extremely conservative) for economic
policies and
$-8$ (extremely liberal) to $+8$ (extremely conservative) for social
policies. Points have been jittered to avoid overplotting. The symbols
in the third graph show the average perceived positions of Bush and
Kerry.}\label{views}
\end{figure}

The analysis presented in the preceding section is interesting,
counterintuitive, and potentially appealing if you think it would be
desirable for the two parties to be further apart, to present a clearer
choice to voters. We test it using voters' placements of themselves and
the candidates on economic and social issues in the 2004 National
Election Study.

We take three questions for each set of issues,
\footnote{The social
issues were opinions about the role of women, gun-control policy, and
government aid to African Americans. The economic issues were opinions
about the level of spending that the government should undertake in the
economy, the role of the government in providing an economic
environment where there is job security, and the level at which the
government should spend on defense. We replicated our analysis removing
the defense spending question (which is arguably on a different
dimension than economics) and got similar results [Cai (\citeyear{Cai2006})].}
using
all the relevant questions from the National Election Study in which
respondents were asked to judge the positions of Bush, Kerry, and
themselves. We then summed the responses in each dimension, yielding a
$-9$ to $9$ scale on economic issues and a $-8$ to $8$ scale on social
issues. We then have six data points for each respondent, representing
economic and social positions as judged for Bush, Kerry, and self.
Figure \ref{views} displays the data: there is correlation across issue
dimensions and also a lot of variation.
It is perhaps surprising that voters differed so much in their
assessments of where Bush and Kerry stand on the issues.

To estimate the effect of a change in party positions, we model in
three steps the data on issue attitudes and vote preference. First, we
fit linear regressions to predict views of Bush's and Kerry's policy
positions, given respondent's party identification and self-placements
on the issues. Second, we fit logistic regressions to the probability
of supporting Bush (among those respondents who express a preference),
given respondent's party identification and his or her relative
distance from each candidate on the issues. Third, we consider
counterfactuals in which the candidates' perceived issue positions
change (by altering the intercepts in the regression in the first stage
of the model), and then seeing the effect in aggregate vote preferences
as predicted by the logistic regressions.

\subsection*{Model of perceived candidate issue positions given self-placements}

\begin{figure}

\includegraphics{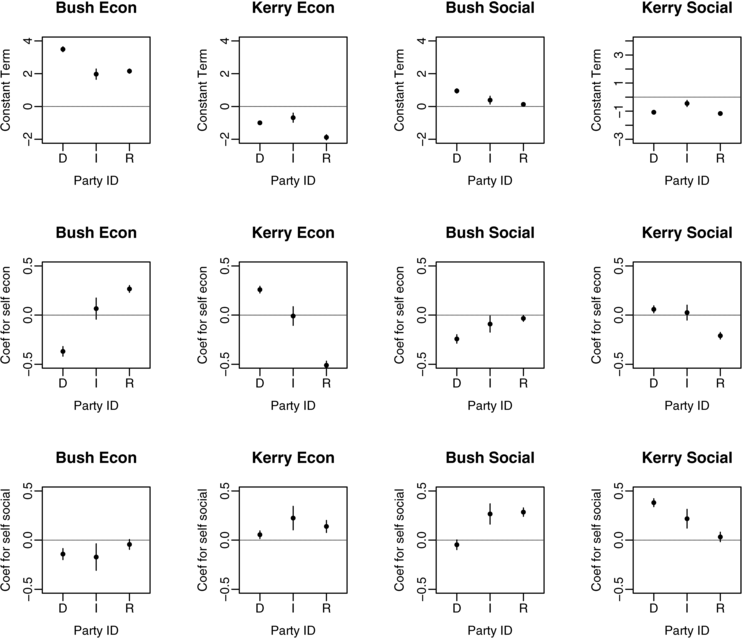}

\caption{Estimated coefficients for the regressions of perceptions of
Bush and Kerry on economic and social issues. The four columns of the
display represent these four outcomes. For each, the top, middle and
bottom rows show estimates ($\pm$1 standard error) of the constant
term and the coefficients for self-perception on economic and social
issues. The model was fit separately to Democrats, independents and
Republicans, as indicated by the three points within each graph.
Generally, Democrats with more liberal positions viewed Kerry as more
liberal and Bush as more conservative on the issues, and Republicans
show the opposite pattern.\label{perceptions}}
\end{figure}

We fit separate regressions on four different outcomes---views of
Bush's and Kerry's position on economic and social issues---and
the display in
Figure \ref{perceptions} shows the estimated coefficients for the
constant term and for self-perceptions on economic and social issues.
Within each of the twelve plots are the estimates for the models fit
separately to Democrats, independents and Republicans.

In considering Figure \ref{perceptions}, we first discuss the two
columns on the left, which relate to views of the candidates' economic
positions.
The constant terms show, unsurprisingly, that Bush is viewed as more
conservative than Kerry, with Democrats perceiving Bush as more
conservative and Republicans perceiving Kerry as more liberal. The
coefficients for self-perception on economic issues show a striking
pattern: the more liberal a Democrat is on economic issues, the more he
or she views Bush as conservative and Kerry as liberal, with the
reverse happening for Republicans. Apparently, there is a strong
motivation to believe that your party's candidate is similar to you in
his political views. Weaker patterns appear in the first two plots in
the lowest row of Figure \ref{perceptions} with self-perceptions on
social issues being slightly negatively predictive of views of Bush's
economic position and slightly positively predictive of views on Kerry.

We now consider the two columns on the right of Figure
\ref{perceptions}. Again, the intercepts are higher for Bush than for
Kerry, but to a much weaker extent than for the economic position,
indicating that more of the variation in views of the candidates'
positions on social issues is explained by respondents'
self-perceptions. Here the patterns are more complex. Democrats' views
of Bush's position on social issues is negatively predicted by
self-perceptions on {\em economic} issues, with self-perception on
social issues not coming into the equation at all. In contrast,
Democrats' views of Kerry's positions on social issues are entirely
predicted by self-perceptions on {\em social} issues.
Now we look at the coefficients for Republican respondents: to predict
their views of Bush's position on social issues, only their
self-perception on social issues is relevant, but when predicting
Kerry's position on social issues, only their self-perception on
economics is relevant.

To summarize, voters appear to characterize their own party's nominee's
positions in a way consistent with their self-perception on each issue
dimension. But their views of the other party's nominee, in both
dimensions, is predicted (with a negative coefficient) solely based on
self-perception on economics.

\subsection*{Model of vote choice given distances from candidates}

Our next step is a logistic regression model predicting vote preference
given ideological distance from candidates. We define, for each survey
respondent $i$, the distance from Bush minus the distance from Kerry:
\begin{eqnarray*}
(\mathrm{dist.E})_i &=& (\mathrm{econ}^{\mathrm{Bush}}_i
- \mathrm{econ}^{\mathrm{self}}_i)^2
- (\mathrm{econ}^{\mathrm{Kerry}}_i
- \mathrm{econ}^{\mathrm{self}}_i)^2,
\\
(\mathrm{dist.S})_i &=& (\mathrm{soc}^{\mathrm{Bush}}_i
- \mathrm{soc}^{\mathrm{self}}_i)^2
- (\mathrm{soc}^{\mathrm{Kerry}}_i
- \mathrm{soc}^{\mathrm{self}}_i)^2,
\end{eqnarray*}
and then we fit a logistic regression of vote intention ($y_i=1$ if
respondent $i$ supports Bush for President, 0 for Kerry, excluding
undecideds and others from the analysis) on
$\mathrm{dist.E}$ and $\mathrm{dist.S}$.
We fit separate models for each party identification, yielding
%
\begin{eqnarray}\label{3models}
\Pr(y_i=1) &=& \mathrm{logit}^{-1}
\bigl(-1.32 -0.05 \cdot(\mathrm{dist.E})_i
-0.04 \cdot(\mathrm{dist.S})_i\bigr)
\nonumber\\
\eqntext{\mbox{ for Democrats},}
\\
\Pr(y_i=1) &=& \mathrm{logit}^{-1}
\bigl(0.38 -0.05 \cdot(\mathrm{dist.E})_i +
0.02 \cdot(\mathrm{dist.S})_i\bigr)
\nonumber\\[-8pt]
\\[-8pt]
\eqntext{\mbox{ for independents},}
\\
\Pr(y_i=1) &=& \mathrm{logit}^{-1}\bigl(2.30 -0.03 \cdot(\mathrm{dist.E})_i -0.02
\cdot(\mathrm{dist.S})_i\bigr)
\nonumber\\
\eqntext{\mbox{ for Republicans.}}
\end{eqnarray}
As expected, $\mathrm{dist.E}$ (economics) is more important than
$\mathrm{dist.S}$
(social issues), and the coefficients themselves are negative: if you
are further from Bush than from Kerry, you are less likely to support
Bush. The only exception is the positive coefficient for
$\mathrm{dist.S}$
among independents, but this is not statistically significant (the
estimate is 0.02 with a standard error of 0.02) so we take it to just
represent sampling error. We also see that the coefficients for
ideological distance are greater for Democrats than for Republicans,
which is consistent with the idea that Democrats are more diverse in
their political preferences (so that conservative Democrats are more
likely to vote for Bush than liberal Republicans were to vote for Kerry).

\subsection*{Model of aggregate vote given shifts in candidates\textup{'} positions}

Our next step is to consider hypothetical changes in the candidates'
positions on economic and social issues, and see how these would
translate into vote changes. For each change, we simply alter the
constant term in the appropriate regressions shown in Figure~\ref{perceptions}---for
example, if we want to shift Kerry by one point to
the right on the $-9$ to $9$ economic scale, we add 1 to the intercepts
of the ``Kerry econ'' regressions for each of the three party
identification groups. We then run the models of the previous sections
forward, first simulating random positions from the linear models from
Figure~\ref{perceptions} (with intercepts altered appropriately), then
computing estimated ideological distances and simulating vote
preferences from the logistic regressions~(\ref{3models}). This
represents a replicated election outcome under the hypothetical
position shift. For each hypothesized shift, we take the average of
100~simulations to get the predicted election outcome.

\begin{figure}

\includegraphics{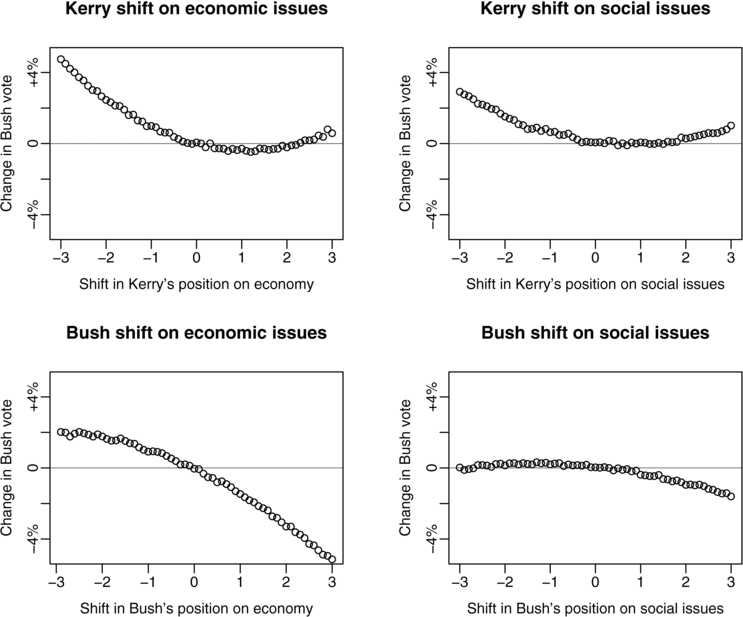}

\caption{Predicted change in Bush\textup{'}s share of the vote, if Kerry\textup{'}s or
Bush\textup{'}s position on economic or social issues were to shift by a
specified amount. The predictions are calculated based on the fitted
logistic model of vote choice given voters\textup{'} ideological distances from
candidates. Positions on the economy and on social issues are measured
on a $-9$ to 9 scale, and a $-8$ to 8 scale, respectively; see Figure
\protect\ref{views}. Based on this model, it would be beneficial for Kerry to
shift slightly to the right in both dimensions, for Bush to shift
slightly to the left on social issues, and for Bush to shift a great
deal to the left on economic issues.
The curves are slightly jittery because of simulation
variability.\label{1d.shifts}}
\end{figure}
\begin{figure}

\includegraphics{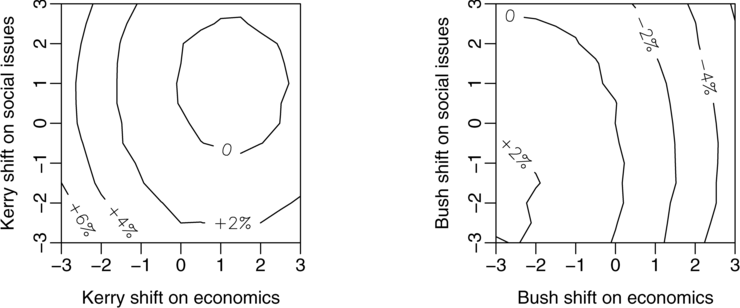}

\caption{Predicted change in Bush\textup{'}s share of the vote, if Kerry\textup{'}s or
Bush\textup{'}s position were to shift on both economic and social issues.
According to this model, the optimal strategy for Kerry is to move~1
point to the right in both dimensions; in contrast, Bush would benefit
by moving about 2 points to the left on social issues and nearly 3
points to the left on the economy.\label{2d.shifts}}
\end{figure}

Figure \ref{1d.shifts} shows the effect, under this model, of shifting
the positions of either Kerry or Bush on economic or social issues, by
as much as 3 points in either direction. The answer to the question
posed by the title of the paper appears to be No, Kerry should not have
moved to the left on economic policy. Conventional wisdom appears to be
correct: Kerry would have benefited by moving to the right, and Bush by
moving to the left. The optimal shifts for Bush are greater than those
for Kerry, which is consistent with the observation that voters are, on
average, closer to the Democrats on issue attitudes.

Figure \ref{2d.shifts} shows similar calculations, allowing each
candidate to move in both issue dimensions. Again, this model finds
Kerry benefiting by moving a bit to the right, and Bush benefiting by
moving a lot to the left, especially in the economic dimension. One
could continue along these lines by allowing the two candidates to move
simultaneously, but this is not our goal here. We do not consider our
calculations to represent a realistic causal model of what would happen
if candidates were to move; rather, it is a way of exploring the
multidimensional space of voters' perceptions of themselves and the
candidates, and evaluating in a fairly direct way the hypotheses of
Section \ref{theory}.

Comparing these shifts to the candidates' average perceived positions
(see the rightmost plot in Figure \ref{views}), the optimal position for Kerry is
to the right of his position at the time, but still far to the left of
the perceived position of Bush. Given that the Republicans are far to
the right of the median voter on economic issues---and given the large
variation in voters' perceptions of the candidates' positions---it
appears to be best for the Democrats to stay in the center, quite a bit
left of the Republicans, in order to make their relative location clear
to the voters.

\subsection*{Comparison to the theoretical model}

The empirical model we have used is a generalization of the formal
model of Section \ref{theory}, in four ways: (a) voters are allowed to
vary in their perceived positions of the candidates, (b) candidates can
differ in their valences, (c) the two issue dimensions need not be
equal in importance, and (d) different models apply to Democrats,
Republicans and independents. The model (\ref{3models}) is equivalent
to a spatial voting model with weighted squared Euclidean distance and
logistic errors (the discrete responses have enough different
categories that the continuous approximation seems reasonable enough).

The empirical conclusions are similar but not identical to the results
of Section \ref{theory}, with the key difference being that the voters
on average perceived Kerry as slightly left of center on economic
issues [see Figure \ref{views}(c)], as compared to the theoretical model
of Figure \ref{2d}(a), which hypothesized that voters saw little
difference between the candidates on this dimension. In both the
theoretical and empirical models, the Democrats would benefit by
placing or maintaining some distance between themselves and the
Republicans on economic issues.

\section{Practical concerns}\label{practical}

\subsection{Using survey responses to measure perceived ideological positions}

\hspace*{-1pt}A~key issue regarding with the empirical part of this study is the
reliability and validity of the survey questions about candidate- and
self-placement. It has long been known that responses to individual
issue positions are unstable over time and are not meaningful for many
voters [Erikson and Tedin (\citeyear{eriTed2004})]. In our data this can be seen in the
wide variation in perceptions of Bush and Kerry on the issues (see the
left two plots in Figure \ref{views}).
Ansolabehere, Rodden and Snyder (\citeyear{ansRodSny2006})
have shown that more can be
learned by averaging the responses to several related questions. Our
economic and social attitude scales are based on only three questions
each (in the National Election Study, all we could find that asked
about the candidates and the respondent), and we would be interested in
results from a more detailed survey. On the other hand, if the goal is
to model what would happen if candidate positions change, this needs to
be filtered through the imperfect perceptions of voters, so it is not a
fatal flaw that respondents are not completely consistent with
themselves and each other.

Another concern is the complicated nonsequential relationship between
party identification, issue attitudes, perceptions of candidates, and
vote preference [Page and Jones (\citeyear{pagJon1979})].
Party identification is a stable individual measure
[Miller and Shanks (\citeyear{milSha1996})],
so we do not mind subdividing our analysis into Democrats,
independents and Republicans. Beyond this, we recognize that it is an
approximation to model vote preference as a function of candidate
perceptions rather than the reverse. Our regressions are based on the
observed correlations between the issue-response and vote-choice
questions, and we are implicitly making additional causal assumptions
in using the model to speculate on what would happen if the candidate
positions changed. We think our approach is a useful starting point,
however, and even this imperfect empirical analysis gives insight into
models such as the median voter theorem that are commonly applied
automatically without any connection to data.

\subsection{Constraints and flexibility in party positions}

Our analysis treats the two parties asymmetrically and treats the issue
dimensions differently as well. Is it reasonable to suppose that the
Republicans cannot move ideologically but the Democrats can? And is it
plausible that the Democrats are free to move to the left on economic
policy but cannot move to the center on foreign policy and social issues?

We would answer Yes to both these questions. It is reasonable to
suppose that, as the party in power, the Republicans are less inclined
to make an ideological move that would convince the voters. In
addition, their conservative position on economic issues is important
to a key segment of the Republicans' electoral, financial, and
intellectual base. It makes sense that the Republicans will remain to
the right of the majority of voters on economic issues, even if this
costs them some votes.

As for the Democrats, we would expect that most of their stakeholders
would prefer a move to the left on economic issues---if anything, it
might be that their moderately conservative position was chosen partly
from a median-voter thinking as exemplified by Figure \ref{1d}.
\footnote{Moving to the left would not be costless for the Democrats, however.
In particular, one would expect them to lose some contributions from
businesses and affluent individuals, and support of policies such as
tariff barriers could be unpopular among elite opinion-makers such as
those who determine newspaper endorsements.}
In contrast, internal
party pressures could make it more difficult for the Democrats to move
toward the center in other dimensions.

A related question is how the parties can signal their position changes
to the voters. Our models simply assume the ability to do so, but
presumably the voters would need some convincing that a move to the
center is not just a pre-election ploy.

Finally, a common counter-argument to spatial voting models is that
moving to the median might gain votes at the middle at the expense of
the other party, but at the cost of diminishing turnout among one's
core supporters. There is no particular evidence that this happened in
2004. The innovation of the theoretical model of this paper is to posit
a counterintuitive motivation for a party (in this case, the Democrats)
to distinguish itself in policy from the other part, purely from
spatial voting concerns arising from multidimensionality and variation
in voters' perceptions of the candidates, \textit{without} bringing in turnout.

\section*{Acknowledgments}

We thank Jasjeet Sekhon, Shigeo Hirano, Robert Erikson, Jeff Lax, Joseph
Bafumi and David Park for helpful conversations.

\printaddresses

\end{document}